\documentclass[twocolumn,pre]{revtex4}

\usepackage{graphicx}      % for figures
\usepackage{amsmath}       % for subequations
\usepackage{amssymb}       % for proof boxes
\usepackage{bm}            % for bold math type
\usepackage{units}         % for vulgar fractions
\usepackage{hyperref}      % for hypertext
\hypersetup{colorlinks,citecolor=black,filecolor=black,linkcolor=black,urlcolor=black}

\begin{document}

\title{Invariant submanifold for series arrays of Josephson junctions}

\author{Seth A. Marvel}
\email{sam255@cornell.edu}

\author{Steven H. Strogatz}
%\email{strogatz@cornell.edu}

\affiliation{Center for Applied Mathematics, Cornell University, Ithaca, New York 14853}

\begin{abstract}
We study the nonlinear dynamics of series arrays of Josephson junctions in the large-$N$ limit, where $N$ is the number of junctions in the array.  The junctions are assumed to be identical, overdamped, driven by a constant bias current and globally coupled through a common load.  Previous simulations of such arrays revealed that their dynamics are remarkably simple, hinting at the presence of some hidden symmetry or other structure.  These observations were later explained by the discovery of $N-3$ constants of motion, each choice of which confines the resulting flow in phase space to a low-dimensional invariant manifold.  Here we show that the dimensionality can be reduced further by restricting attention to a special family of states recently identified by Ott and Antonsen.  In geometric terms, the Ott-Antonsen ansatz corresponds to an invariant submanifold of dimension one less than that found earlier.  We derive and analyze the flow on this submanifold for two special cases:  an array with purely resistive loading and another with resistive-inductive-capacitive loading.  Our results recover (and in some instances improve) earlier findings based on linearization arguments.  \end{abstract}

\maketitle

\textbf{Josephson junctions are superconducting devices with many practical applications, ranging from voltage standards to ultrasensitive detectors.  From a mathematical perspective, their nonlinear dynamics are fascinating, especially when many junctions are coupled together in an array.  For about the past twenty years, theorists have been intrigued by the strange collective behavior seen in numerical experiments on arrays of identical junctions in series.  The behavior began to make sense when it was eventually realized that despite the presence of dissipation in the underlying circuits, the equations possess an enormous number of constants of motion.  These constants restrict the dynamics to low-dimensional manifolds in phase space.  In this paper, we show that in certain cases the reduced dimensionality can be even more severe than previously realized.  By making use of an ansatz recently introduced by Ott and Antonsen, we demonstrate for example that a resistively loaded array can behave exactly as if its phase space were two-dimensional, even when the array consists of infinitely many junctions.}

\section{INTRODUCTION \label{INTRODUCTION}}

Forty years ago, Winfree pioneered the study of synchronization in large populations of coupled limit-cycle oscillators~\cite{winf67}.  Since then, the field has expanded considerably, thanks in large part to Kuramoto's elegant reformulation~\cite{kura84,aceb05} of Winfree's intuitive model.  Both of their models were originally motivated by biological phenomena~\cite{stro94} such as the alpha rhythm of brain waves~\cite{wien58,wien61,fran00}, the collective firing of cardiac pacemaker cells~\cite{pesk75,mich86,mich87}, the coordinated flashing of southeast Asian fireflies~\cite{buck68,hans78,buck88,erme91,kimd04}, menstrual synchrony among close female friends~\cite{mccl71,russ80,wils92}, and glycolytic oscillations in yeast populations~\cite{ghos71,njus81,hast85,dano01,murr01,ojal04}.
However, the techniques developed in analyzing these systems soon proved relevant to physical problems, including the dynamics of laser arrays~\cite{wiwa88,wies90,lier92,fabi93,kour95,kozy00,heil01}, charge-density waves~\cite{stro88,stro89,midd92} and coherence among sites of electrochemical dissolution~\cite{wang00,kiss02,kisz02}.

For simplicity, all the individual oscillators in such models have often been assumed to be coupled equally strongly to all the others, a form of interaction known variously as global, infinite-range, or mean-field coupling.  While this form of coupling is a crude approximation in most cases, series arrays of Josephson junctions constitute a notable exception.  In these systems, exact global coupling between the Josephson junctions emerges naturally from Kirchhoff's laws and the physical properties of weakly coupled superconductors~\cite{tsan91,stro93}.

In the early 1990s, numerical simulations of Josephson junction arrays revealed that they were prone to a large degree of neutral stability~\cite{tsan91,tsan92,nich92,golo92,stro93}.  This peculiar phenomenon was first seen in a simple array in which $N$ identical junctions were connected in series and coupled by a resistor in parallel with all of them.  The numerics suggested that the system's trajectories were always trapped on two-dimensional tori, no matter how many junctions were included in the array~\cite{tsan91}.  Later studies showed that arrays of identical junctions coupled through other kinds of loads displayed a similarly non-generic form of behavior: all but four of the Floquet multipliers for a certain periodic state known as a splay state appeared to lie on the complex unit circle and remain there as the system parameters were varied~\cite{nich92, stro93}.  

These and other puzzling observations were partially explained by the subsequent discovery that the equations of motion could be reduced to a dynamical system of much lower dimension~\cite{wata93,wata94,goeb95}.  Specifically, Watanabe and Strogatz~\cite{wata93,wata94} found a time-dependent trigonometric transformation that expressed the junction phases in terms of $N$ constant phases and three collective time-dependent variables, each obeying a suitable differential equation.  The transformation took the form
	\begin{equation} \label{WS}
	\tan\left[\frac{\phi_j(t)-\Phi(t)}{2}\right] = \sqrt{\frac{1+\gamma(t)}{1-\gamma(t)}} \tan\left[\frac{\theta_j-\Theta(t)}{2}\right]
	\end{equation}
for Josephson junctions $j = 1, \ldots, N$.  Here, the variables $\phi_j(t)$ denote the junction phases, while the constants $\theta_j$ denote the fixed phases on which the transformation operates, and $\Phi(t)$, $\gamma(t)$ and $\Theta(t)$ denote the collective variables.

In the limit of infinitely many Josephson junctions, another result could be extracted from (\ref{WS}).  The transformation maps old phases $\theta_j$ to new phases $\phi_j$ in such a way that a uniform distribution in $\theta$, with phases spread evenly around the circle, is transformed into a non-uniform distribution in $\phi$, with phases symmetrically clumped about some mean phase and distributed according to a Poisson kernel.  This result implies that the space of Poisson kernels is dynamically invariant.  The reasoning is as follows: if the phase distribution takes the form of a Poisson kernel at any time, it can be viewed as having arisen (via the transformation) from an initially uniform phase distribution and hence will remain distributed as a Poisson kernel for all time.

Recently, Ott and Antonsen~\cite{otta08} showed by explicit calculation that the submanifold of Poisson-kernel phase distributions---henceforth called the \textit{Poisson submanifold}---is invariant for a much wider class of models, including the Kuramoto model and other systems in which the oscillators are non-identical.  For the Kuramoto model in which the oscillators have Lorentzian-distributed natural frequencies, Ott and Antonsen~\cite{otta08} derived an exact differential equation for the evolution of the complex order parameter (the centroid of the phase distribution around the unit circle).  As it happens, the amplitude and phase of the order parameter completely characterize the Poisson kernel.  Hence, by extracting the dynamics of the order parameter, Ott and Antonsen~\cite{otta08} simultaneously unveiled the dynamics on the Poisson submanifold.

Our goal in this paper is to use Ott and Antonsen's ansatz~\cite{otta08} to investigate the dynamics of series arrays of identical Josephson junctions.  Before turning to this specific task, however, we first consider the scope of the ansatz itself.  What algebraic form do the governing equations need to have in order for the ansatz to work?  In Sec.~\ref{REDUCIBLE SYSTEMS}, we pinpoint the family of equations that can be simplified in this way, a family that includes Josephson junction series arrays as a special case.

Sections~\ref{RESISTIVELY-LOADED CIRCUIT} and~\ref{RLC-LOADED CIRCUIT} then apply the ansatz to two particular arrays, one with a resistive load and another with an $RLC$ load.  The analysis illuminates the order parameter dynamics on the entire Poisson submanifold.  In this way, we clarify for the first time how the associated arrays behave, not just near their equilibria and periodic orbits, but also far from those special states. 

Finally, Sec.~\ref{DISCUSSION} discusses what the ansatz does---and does not---imply about the dynamics of the original Josephson junction arrays.  Although the ansatz provides powerful insights, it does not tell the whole story; it misses certain important dynamical states that lie off the submanifold of Poisson kernels.  These more general states can be handled by considering (\ref{WS}) within the formalism of M\"obius transformations, as has recently been shown elsewhere~\cite{piko08,miro09}.

\section{REDUCIBLE SYSTEMS \label{REDUCIBLE SYSTEMS}}

The most extensively studied systems of phase oscillators, from the Kuramoto model to Josephson junction arrays, involve purely sinusoidal interactions.  This single-harmonic structure is the key to the success of the Ott-Antonsen ansatz~\cite{otta08}, as the following calculations show.

Consider a system of $N$ identical phase oscillators governed by 
	\begin{equation} \label{gov}
	\dot{\phi_j} = f e^{-i\phi_j} + g + \bar{f}e^{i\phi_j} 
	\end{equation}
for $j = 1, \ldots, N$.  Here $f$ is any smooth, complex-valued, $2 \pi$-periodic function of the phases $\phi_1, \ldots, \phi_N$.  The function $f$ is allowed to depend on time and any other auxiliary state variables in the system (for example, the charge on a load capacitor or the current through a load resistor for the Josephson junction arrays discussed in Sec.~\ref{RLC-LOADED CIRCUIT}).  What is crucial, however, is that $f$ must \emph{not} depend on the oscillator index $j$; it must be the same function for all $j$.  Likewise, the function $g$ must be independent of $j$.  Note that $g$ has to be real-valued since $\dot{\phi_j}$ is real.

Intuitively, the functions $f$ and $g$ can be regarded as common fields felt by all the oscillators.  These fields might involve averages over all the phases, as in models with mean-field coupling, but this is not necessary.  In fact, (\ref{gov}) need not even have permutation symmetry.  That is, the equations need not stay the same under an arbitrary interchange of indices, because the functions $f$ and $g$ need not respect such a symmetry.  The only requirement is that $f$ and $g$ must be the same for all $j$.

The class of systems (\ref{gov}) was studied previously by Watanabe and Strogatz~\cite{wata94} and by Goebel~\cite{goeb95}, who showed that all equations of this form are solved by the transformation (\ref{WS}), where the evolution of $\Phi$, $\gamma$ and $\Theta$ is governed by the forms of $f$ and $g$.  We now show that the newly discovered Ott-Antonsen ansatz also works on this same class of systems, suggesting some intimate relationship between the two reduction methods, as addressed by~\cite{piko08,miro09}.

The Ott-Antonsen ansatz is restricted to the infinite-$N$ limit of (\ref{gov}).  In this limit, one describes the system not in terms of the motion of individual oscillators, but rather in terms of the evolution of the phase density $\rho(\phi,t)$, defined such that $\rho(\phi,t) \mathrm{d}\phi$ gives the fraction of phases that lie between $\phi$ and $\phi + \mathrm{d}\phi$ at time $t$.  Then $\rho$ satisfies the continuity equation 
	\begin{equation} \label{continuity}
	\dot{\rho} + \frac{\partial (\rho v)}{\partial \phi} = 0
	\end{equation}
where the velocity field is
	\begin{equation} \label{velocity}
	v(\phi, t) = f e^{-i\phi} + g + \bar{f}e^{i\phi},
	\end{equation}
from (\ref{gov}).  Here, in the case of infinite $N$, our assumptions about the coefficient functions $f$ and $g$ take the form that $f$ and $g$ may depend on $t$ but not $\phi$.  The time-dependence of $f$ and $g$ can arise either explicitly (through external forcing, say) or implicitly (through the time-dependence of the harmonics of $\rho$ or any auxiliary state variables in the system).

Next, following Ott and Antonsen~\cite{otta08}, suppose $\rho$ is of the form
	\begin{equation} \label{5}
	\rho(\phi, t) = \frac{1}{2\pi} \biggl\lbrace 1+\sum_{n=1}^\infty \bigl(\alpha(t)^ne^{in\phi}+\bar{\alpha}(t)^ne^{-in\phi}\bigr) \biggr\rbrace
	\end{equation}
for some unknown function $\alpha$ that is independent of $\phi$.  Note that (\ref{5}) is just an algebraic rearrangement of the usual form for the Poisson kernel:
	\begin{equation} \label{6}
	\rho = \frac{1}{2\pi} \frac{1-r^2}{1-2r\cos(\phi-\psi)+r^2}
	\end{equation}
where $r$ and $\psi$ are defined via 
	\begin{equation} \label{alpha}
	\alpha = re^{-i\psi}.  
	\end{equation}
In geometrical terms, the ansatz (\ref{5}) defines a submanifold in the infinite-dimensional space of density functions $\rho$.  This submanifold is two-dimensional and is parametrized by the complex number $\alpha$ (or equivalently, by the polar coordinates $r$ and $\psi$).

We now show that the submanifold of Poisson kernels is invariant and calculate the flow on it.  To do so, we substitute the velocity field (\ref{velocity}) and the ansatz (\ref{5}) into the continuity equation (\ref{continuity}).  After reindexing where appropriate, we obtain
	\begin{equation} \label{8}
	\bigl[\dot{\alpha} + i\bigl(\bar{f}+g\alpha+f\alpha^2\bigr)\bigl] \sum_{n=1}^\infty n\alpha^{n-1}e^{in\phi} + \text{c.c.} = 0
	\end{equation}
where c.c.~denotes the complex conjugate of the first term in (\ref{8}).  Note the seemingly miraculous coincidence here: the expression in brackets is a common factor for each term in the sum.  Hence, although (\ref{8}) constitutes an infinite set of amplitude equations, with one for each harmonic, all of them are satisfied simultaneously if the bracketed expression vanishes.  This condition is also necessary, since
	\begin{equation} \label{9}
	\alpha^{-1} \sum_{n=1}^\infty n(\alpha e^{i\phi})^n = \frac{e^{i\phi}}{(1-\alpha e^{i\phi})^2} \neq 0.
	\end{equation}
Thus (\ref{8}) is satisfied for all $\phi$ if and only if
	\begin{equation} \label{10}
	\dot{\alpha} + i\bigl(\bar{f}+g\alpha+f\alpha^2\bigr) = 0.
	\end{equation}

It proves convenient to reexpress this result in terms of the complex order parameter $z$, defined as usual by the centroid of the phase distribution: 
	\begin{equation} \label{z}
	z(t) = \int_0^{2\pi} e^{i\phi} \rho(\phi, t) \mathrm{d}\phi.
	\end{equation}
Then by substituting (\ref{5}) into (\ref{z}) we find that $z = \bar{\alpha} = re^{i\psi}$.  Hence, $z$ satisfies the Riccati equation 
	\begin{equation} \label{riccati}
	\dot{z} = i(f + gz + \bar{f}z^2).
	\end{equation}
This equation gives the flow on the Poisson submanifold.  When $f$ and $g$ are functions of $z$ and $t$ alone, as in the case that they can be expressed in terms of the harmonics of $\rho$, (\ref{riccati}) constitutes a closed two-dimensional system.

	\begin{figure} 
	\centering
	\includegraphics[scale = 1]{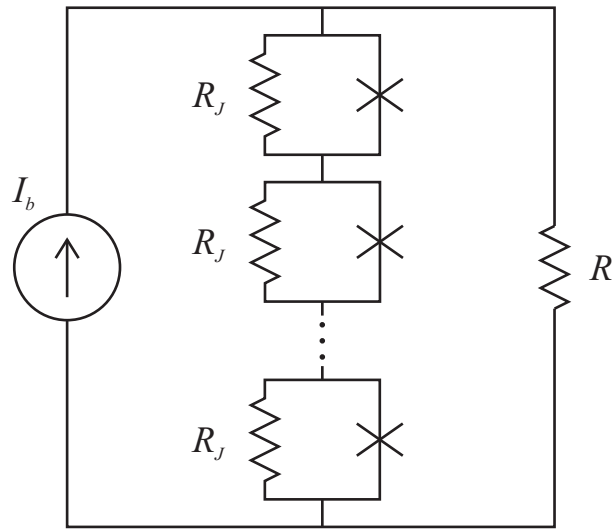}
	\caption{A series of Josephson junctions in parallel with a resistive load.  $I_b$ is the constant bias current, $R$ is the load resistance, and $R_J$ is the internal resistance of a single Josephson junction. \label{rcircuit}}
	\end{figure}

\section{RESISTIVELY-LOADED CIRCUIT \label{RESISTIVELY-LOADED CIRCUIT}}

We now apply (\ref{riccati}) to obtain the reduced dynamics of the circuit shown in Fig.~\ref{rcircuit}.  This circuit consists of $N$ Josephson junctions wired in series and placed in parallel with a resistive load $R$ and a constant current source $I_b$.  Let $I_c$ denote the critical current of each junction and let $R_J$ denote its internal resistance.  For simplicity, the junctions are assumed to be heavily overdamped so that we can neglect their internal capacitance.

As demonstrated by~\textcite{tsan91}, the time-dependent dynamics of this circuit can be converted via Kirchhoff's laws and the physical properties of Josephson junctions to the dimensionless system:
	\begin{equation} \label{11}
	\dot{\phi}_j = \Omega + a\cos\phi_j + \frac{1}{N} \sum_{k=1}^N \cos{\phi_k}
	\end{equation}
for $j = 1, \ldots, N$.  Here, $\phi_j = \delta_j - \pi/2$, where $\delta_j$ is the phase difference across the $j$th oscillating Josephson junction.  We write the system in these unusual variables to highlight its reversibility symmetry; the equations stay the same if we change $\phi_j \rightarrow -\phi_j$ and $t \rightarrow -t$.  The dimensionless groups $a$ and $\Omega$ in (\ref{11}) are given in terms of the original circuit parameters by $a = -R/(N R_J) - 1$ and $\Omega = I_b R/(N I_c R_J)$.  Taking the limit $N \rightarrow \infty$ of (\ref{11}), we obtain the velocity field
	\begin{equation} \label{12}
	v(\phi, t) = \frac{a}{2}e^{-i\phi} + (\Omega + r\cos{\psi}) + \frac{a}{2}e^{i\phi}
	\end{equation}
where $r$ and $\psi$ denote the amplitude and phase of the complex order parameter $z$, as before.  Equation (\ref{12}) has the form of (\ref{velocity}), so by (\ref{riccati}), the order parameter dynamics are
	\begin{equation} \label{13}
	\dot{z} = i\biggl[\frac{a}{2} + (\Omega + r\cos{\psi})z + \frac{a}{2}z^2 \biggr].
	\end{equation}
Taking the real and imaginary parts of (\ref{13}) yields the two-dimensional nonlinear system:
	\begin{equation} \label{14}
	\begin{split}
	\dot{r}    & = \frac{1+b}{2}(r^2-1)\sin{\psi} \\
	\dot{\psi} & = \Omega - \frac{1+b}{2}(r+r^{-1})\cos{\psi} + r\cos{\psi}
	\end{split}
	\end{equation}
where we have introduced the parameter $b = -a-1 = R/(N R_J)$, which simplifies the parameter space and fixed points of (\ref{14}).  Note that $b > 0 $ for any real circuit, although in what follows we will also allow negative values of $b$, since our main interest is with (\ref{11}) as a dynamical system, not as a model of a real device.

\subsection{Fixed points \label{Fixed points 1}}

This section analyzes the fixed points of (\ref{14}) with respect to their dependence on the parameters $b$ and $\Omega$.  There are many cases so we organize the calculations as a series of simple claims.  Readers who prefer to bypass the algebra can skip ahead to Sec.~\ref{Parameter space and phase portraits} which summarizes the findings.

First we convert (\ref{14}) to Cartesian coordinates.  This removes the coordinate singularity at $r=0$ and permits simpler proofs of where the fixed points of (\ref{14}) exist in the $b$-$\Omega$ parameter space.  Let $x = r\cos{\psi}$ and $y = r\sin{\psi}$.  Then (\ref{14}) becomes
	\begin{equation} \label{15}
	\begin{split}
	\dot{x} & = bxy - \Omega y \\
	\dot{y} & = \frac{1-b}{2}x^2 + \Omega x + \frac{1+b}{2}y^2 - \frac{1+b}{2}
	\end{split}
	\end{equation}
Note that (\ref{15}) remains identical under the transformation $\Omega \rightarrow -\Omega$, $x \rightarrow -x$, which represents the symmetry of the resistively-loaded circuit in Fig.~\ref{rcircuit} under a sign reversal of $I_b$ and reflection of the coordinate system on which phase synchrony is measured.  Hence, the number and stability of the fixed points remains unchanged for each point in $b$-$\Omega$ space reflected across the $b$-axis.  Without loss of generality, we therefore consider only positive values of $\Omega$ from here on.

If $b \neq 0, \pm 1$, (\ref{15}) has four fixed points:
	\begin{subequations} \label{16}
	\begin{align}
	x^* & = \Omega/b;\; y^* =  \sqrt{1-\Omega^2/b^2}             \label{16a} \\
	x^* & = \Omega/b;\; y^* = -\sqrt{1-\Omega^2/b^2}             \label{16b} \\
	x^* & = \frac{-\Omega+\sqrt{\Omega^2-b^2+1}}{1-b};\; y^* = 0 \label{16c} \\
	x^* & = \frac{-\Omega-\sqrt{\Omega^2-b^2+1}}{1-b};\; y^* = 0 \label{16d}	
	\end{align}
	\end{subequations}
The points (\ref{16a}) and (\ref{16b}) lie on the boundary $r=1$ of the unit disk and represent \emph{synchronized rest states}: equilibrium states in which all the Josephson junctions are perfectly in phase and do not oscillate.  In this case, all the individual phases $\phi_j$ equal the same constant, and hence equal the constant phase $\psi$ of the centroid.  Physically, such states would be superconducting.  The source current tunnels through each of the $N$ junctions without developing any voltage across the load.

In contrast, the fixed points (\ref{16c}) and (\ref{16d}) have $r < 1$ and lie inside the unit disk, meaning that the Josephson junctions are not all in phase.  This type of fixed point is known as a \emph{splay state}~\cite{tsan91,tsan92,nich92,golo92,stro93}.  It represents a periodic collective state in which the junctions oscillate out of phase, but in such a highly organized way that the overall phase distribution remains stationary. In particular, the macroscopic order parameters $r$ and $\psi$ stay constant even though individual junctions change their state non-uniformly, hesitating at some phases and accelerating at others.  The stationary distribution of phases takes the form of a Poisson kernel, as expected.

\subsection{Partitioning the parameter space \label{Partitioning the parameter space}}

We now determine where in the $b$-$\Omega$ parameter space the fixed points exist.  By inspection, the synchronized rest states (\ref{16a}) and (\ref{16b}) exist if and only if $\Omega \leq |b|$.  Meanwhile, for the splay states (\ref{16c}) and (\ref{16d}) to exist, $x^*$ must be real (i.e. $\Omega^2-b^2+1 \geq 0$) and $|x^*| \leq 1$.  The condition $|x^*| \leq 1$ places additional restrictions on the $b$ and $\Omega$ values at which (\ref{16c}) and (\ref{16d}) exist.  We now derive these restrictions explicitly, with the key results summarized in Table \ref{conditions}. \\

\textbf{Claim:} On $\Omega,b > 0$, (\ref{16c}) exists if and only if $\Omega \geq b$.

\textbf{Proof:} Let $h(b,\Omega) = \Omega^2-b^2+1$.  Then by algebraic rearrangement,
	\begin{equation} \label{17}
	\left(\frac{-\Omega+\sqrt{h}}{1-b}\right)^2 \leq 1 \Leftrightarrow \Omega\sqrt{h} \geq \Omega^2-b^2+b.
	\end{equation}

	\begin{table}	\caption{Existence of the fixed points for $\Omega > 0$. \label{conditions}}
		\centering
		\begin{tabular*}{0.37\textwidth}{@{\extracolsep{\fill}} c c}
		\toprule
		\textit{fixed point}    & \textit{conditions}                          \\
		\colrule
		(\ref{16a}),(\ref{16b}) & $\Omega \leq |b|$                            \\
		(\ref{16c})             & $\Omega^2-b^2+1 \geq 0$ and $\Omega \geq b$  \\
		(\ref{16d})             & $\Omega^2-b^2+1 \geq 0$ and $\Omega \leq -b$ \\
		\botrule
		\end{tabular*}
	\end{table}	
	\begin{ruledtabular}
	\begin{table}[!ht]
	\caption{Linearization of (\ref{15}) at (\ref{16a}) for $\Omega > 0$. \label{linear1}}
		\begin{tabular}{c c c}
		\textit{interval of b} & \textit{signs of $\Delta$, $\tau_{+}$}  & \textit{stability}      \\
		                       &                                         & \textit{classification} \\
		\colrule
		$(-\infty, \min\{-1,-\Omega\})$          & $\Delta > 0; \tau_{+} < 0$ & stable node   \\	
		$(-1, \min\{-\nicefrac{1}{2},-\Omega\})$ & $\Delta < 0; \tau_{+} < 0$ & saddle point  \\	
		$(-\nicefrac{1}{2}, -\Omega)$            & $\Delta < 0; \tau_{+} > 0$ & saddle point  \\	
		$(\Omega, +\infty)$                      & $\Delta > 0; \tau_{+} > 0$ & unstable node \\				
		\end{tabular}
	\end{table}
	\end{ruledtabular}	
	
	\begin{ruledtabular}
	\begin{table}[!ht]
	\caption{Linearization of (\ref{15}) at (\ref{16c}), (\ref{16d}) for $\Omega > 0$. \label{linear2}}
		\begin{tabular}{c c c c}
		\textit{fixed point} & \textit{interval of b} & \textit{sign of $\Delta_\pm$} & \textit{stability} \\
                         &                        &                               & \textit{classification} \\
		\colrule
		(\ref{16c}) & $(-\sqrt{\Omega^2+1}, \Omega)$  & $\Delta_{+} > 0$ & center       \\
		\colrule
		(\ref{16d}) & $(-\sqrt{\Omega^2+1}, \qquad$   & $\Delta_{-} < 0$ & saddle point \\
		            & $\qquad \min\{-1,-\Omega\})$    &                  &              \\		
                & $(-1, -\Omega)$                 & $\Delta_{-} > 0$ & center       \\
		\end{tabular}
	\end{table}
	\end{ruledtabular}

When the discriminant of (\ref{16c}) and (\ref{16d}) is nonnegative, $\Omega^2-b^2+b \geq b-1$.  Furthermore, $b-1 \geq 0$ for $b \geq 1$, and $b(1-b) \geq 0$ for $b$ on $[0,1]$, so both sides of the second inequality in (\ref{17}) are nonnegative for all $b > 0$.  Thus, we can square both sides of this inequality to remove the square root:
	\begin{equation} \label{18}
	\Omega^2 h \geq (\Omega^2-b^2+b)^2 \Leftrightarrow \Omega \geq |b|.
	\end{equation}
We can also see that (\ref{18}) implies (\ref{17}) when $h \geq 0$, because there the larger quantity of the first inequality in (\ref{18}) is nonnegative and remains so upon removing the square in going from (\ref{18}) to (\ref{17}).  Hence, the claim is proved. $\square$ \\

\textbf{Claim:} On $\Omega,-b > 0$, (\ref{16c}) exists if and only if $h \geq 0$.

\textbf{Proof:} In the forward direction, we again obtain (\ref{17}), which we split into two cases:
	\begin{subequations} \label{19}
	\begin{align}
	\Omega\sqrt{h} \geq |\Omega^2-b^2+b| \Leftrightarrow \Omega \geq -b, \label{19a} \\
	\Omega\sqrt{h} \leq |\Omega^2-b^2+b| \Leftrightarrow \Omega \leq -b. \label{19b}
	\end{align}
	\end{subequations}
Since regions of $\Omega \geq -b$, $\Omega \leq -b$ both exist on $\Omega,-b,h \geq 0$, this direction does not yield any additional restrictions.  In the reverse direction, we can drop the absolute value signs from (\ref{19a}) since the larger quantity remains positive when $h \geq 0$.  (\ref{19b}) also implies (\ref{17}) if $\Omega^2-b^2+b \leq 0$.  Since $\Omega \leq -b$ implies $\Omega^2-b^2+b \leq 0$ for $b < 0$, (\ref{16c}) exists everywhere its discriminant is nonnegative (on the quadrant $\Omega,-b > 0$). $\square$ \\

\textbf{Claim:} On $\Omega > 0$, (\ref{16d}) exists if and only if $h \geq 0$ and $\Omega \leq -b$.

	\begin{figure}[!ht] 
	\includegraphics[scale = 1]{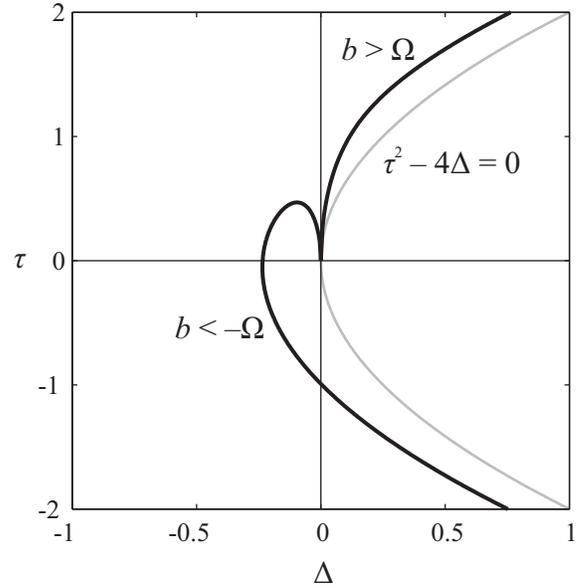}
	\caption{$\Delta$ and $\tau$ of the linearization of (\ref{16a}) plotted parametrically as a function %
		of $b$ for $\Omega = \nicefrac{1}{8}$. \label{tracevsdet}}
	\end{figure}

\textbf{Proof:} Again by algebra,
	\begin{equation} \label{20}
	\left(\frac{-\Omega-\sqrt{h}}{1-b}\right)^2 \leq 1 \Leftrightarrow \Omega\sqrt{h} \leq -\Omega^2+b^2-b.
	\end{equation}
We can square both sides of (\ref{20}) to obtain:
	\begin{equation} \label{21}
	\Omega^2 h \leq (\Omega^2-b^2+b)^2 \Leftrightarrow \Omega \leq |b|.
	\end{equation}
In the reverse direction, removing the square from the first inequality of (\ref{21}) requires that $\Omega^2-b^2+b \leq 0$.  However, $b(1-b) \geq 0$ for $b$ on $[0,1]$ and $\Omega^2-b^2+b \leq 0$ implies $h < 0$ for $b > 1$, so this inequality is \emph{not} satisfied for $b > 0$.  Nevertheless, $\Omega \leq -b$ implies $\Omega^2-b^2+b \leq 0$ for $b < 0$, so (\ref{16d}) exists if and only if $h \geq 0$ and $\Omega \leq -b$. $\square$

\subsection{Stability of the fixed points \label{Stability of the fixed points}}

At (\ref{16a}) and (\ref{16b}), the linearization of (\ref{15}) has the determinant and trace:
	\begin{equation} \label{22}
	\begin{split}
	\Delta   & = b(b+1)(1-\Omega^2/b^2), \\
	\tau_\pm & = \pm(2b+1)\sqrt{1-\Omega^2/b^2}.
	\end{split}
	\end{equation}
where (\ref{16a}) has trace $\tau_{+}$ and (\ref{16b}) has trace $\tau_{-}$.  By careful consideration of (\ref{22}) and Fig.~\ref{tracevsdet}, we find that $\Delta$ and $\tau_{+}$ take signs according to the four cases in Table \ref{linear1}.  The case of $\Delta$ and $\tau_{-}$ is analogous.

Similarly, the determinant and trace of the linearization at (\ref{16c}) and (\ref{16d}) are
	\begin{subequations} \label{23}
	\begin{align}
	\Delta_\pm & = \pm\frac{\Omega}{1-b}\sqrt{h} - \frac{b}{1-b}h, \label{23a} \\
	\tau       & = 0.                                              \label{23b}
	\end{align}
	\end{subequations}
where (\ref{16c}) has determinant $\Delta_{+}$ and (\ref{16d}) has determinant $\Delta_{-}$.  We now consider how $\Delta_\pm$ in (\ref{23a}) takes signs as a function of $b$ and $\Omega$.  The results are summarized in Table \ref{linear2}. \\

\textbf{Claim:} On $\Omega > 0$, (\ref{16c}) is a center. 

\textbf{Proof:} Clearly, $\Delta_{+} > 0$ for $b < 0$.  On $b > 1$,
	\begin{equation} \label{24}
	\Delta_{+} > 0 \Leftrightarrow \frac{|b|}{|1-b|}h > \frac{\Omega}{|1-b|}\sqrt{h} \Leftrightarrow \Omega > b.
	\end{equation}
Likewise, for $b$ on $(0,1)$,
	\begin{equation} \label{25}
	\Delta_{+} > 0 \Leftrightarrow \frac{\Omega}{|1-b|}\sqrt{h} > \frac{|b|}{|1-b|}h \Leftrightarrow \Omega > b.
	\end{equation}
Since $\Omega \geq b$ for all $(b,\Omega)$ where (\ref{16c}) exists, $\Delta_{+} > 0$ and (\ref{16c}) is a center on $b > 0$, as well. $\square$ \\ 

\textbf{Claim:} On $\Omega > 0$, (\ref{16d}) is a center for $b > -1$ and a saddle for $b < -1$. 

\textbf{Proof:} We need only be concerned with the negative $b$-axis, since (\ref{16d}) does not exist where $\Omega,b > 0$.  For $b$ on $(-1,0)$,
	\begin{equation} \label{26}
	\Delta_{-} > 0 \Leftrightarrow \frac{|b|}{|1-b|}h > \frac{\Omega}{|1-b|}\sqrt{h} \Leftrightarrow \Omega < -b,
	\end{equation}
while for $b < -1$,
	\begin{equation} \label{27}
	\Delta_{-} < 0 \Leftrightarrow \frac{\Omega}{|1-b|}\sqrt{h} > \frac{|b|}{|1-b|}h \Leftrightarrow \Omega < -b.
	\end{equation}
Since $\Omega \leq -b$ for all $(b,\Omega)$ where (\ref{16d}) exists, $\Delta_{-} > 0$ and (\ref{16d}) is a center for $b$ on $(-1,0)$, while $\Delta_{-} < 0$ and (\ref{16d}) is a saddle for $b < -1$. $\square$ \\

\subsection{Parameter space and phase portraits \label{Parameter space and phase portraits}}

\begin{figure*}
	\includegraphics[scale = 1]{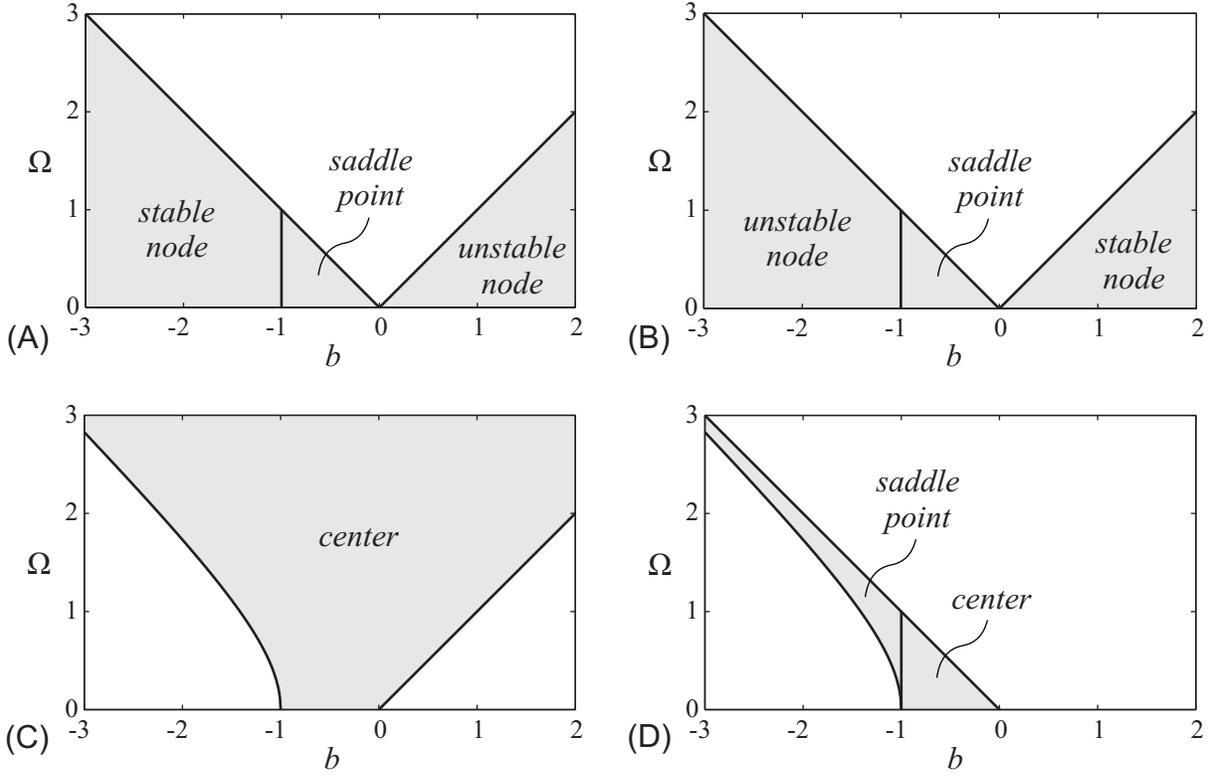}
	\caption{Regions of existence in the upper-half $b$-$\Omega$ plane for each of the four fixed points.  (A) corresponds to fixed point (\ref{16a}), (B) to fixed point (\ref{16b}), (C) to (\ref{16c}) and (D) to (\ref{16d}).  The existence and stability of (\ref{16a}) at a given point $(b,-\Omega)$ of parameter space for $\Omega > 0$ is given by the existence and stability of (\ref{16a}) at $(b,\Omega)$, and likewise for (\ref{16b}).  By contrast, the existence and stability of (\ref{16c}) at $(b,-\Omega)$ is equivalent to the existence and stability of (\ref{16d}) at $(b,\Omega)$, and vice versa. \label{existregions}}
	\end{figure*}

Figure~\ref{existregions} summarizes our findings regarding the regions of the $b$-$\Omega$ parameter plane where each fixed point exists, as well as the stability classifications of the fixed points on these regions.  If we let $\Omega \rightarrow -\Omega$, then $|x^*|$ of (\ref{16c}) becomes $|x^*|$ of (\ref{16d}) and vice versa.  Hence, the existence and stability of (\ref{16c}) at a given point $(b,-\Omega)$ of parameter space for $\Omega > 0$ is given by the existence and stability of (\ref{16d}) at $(b,\Omega)$, and vice versa.

\begin{figure}
	\centering
	\includegraphics[scale = 1]{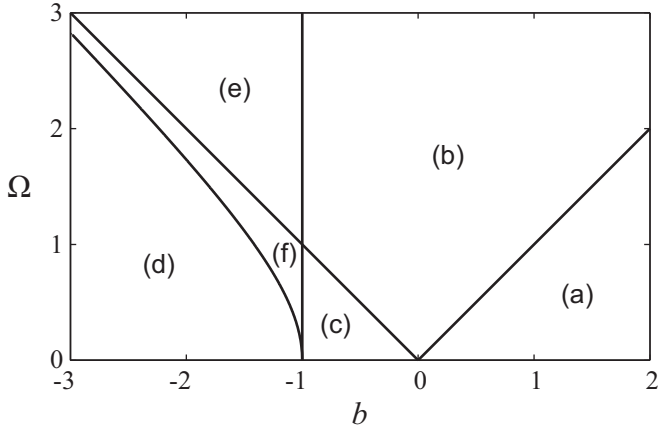}
	\caption{The six qualitatively distinct regions of the $b$-$\Omega$ parameter plane.  The partition is symmetric about the $b$-axis. \label{paramplane}}
	\end{figure}

Figure~\ref{paramplane} combines the four separate panels of Fig.~\ref{existregions} into a single image.  The various regions in Fig.~\ref{existregions} yield six qualitatively distinct regions of the $b$-$\Omega$ plane in Fig.~\ref{paramplane}.  We distinguish between the regions (b) and (e) in Fig.~\ref{paramplane}, because the $x^*$ at which the single center of these regions is located changes sign upon crossing $b = -1$.

\begin{figure*}
	\includegraphics[scale = 1]{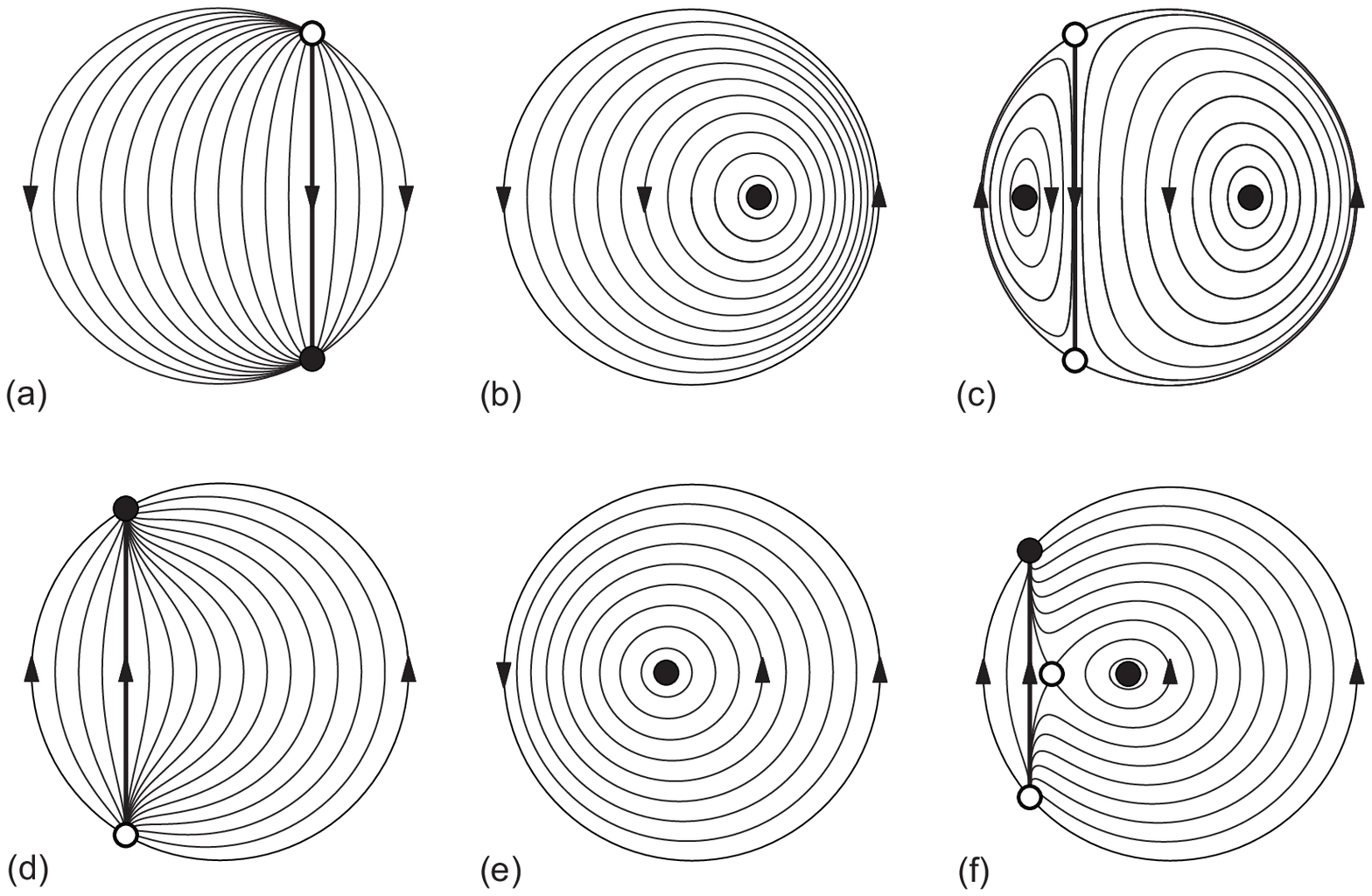}
	\caption{Representative phase portraits for the six regions of the upper-half $b$-$\Omega$ plane with qualitatively distinct phase plane behavior.  The trajectories are plotted in polar coordinates $r$ and $\psi$ on the unit disk.  Solid dots denote Lyapunov stable fixed points, while open dots denote unstable
fixed points.  The letter labels (a)-(f) match with the labels in Fig.~\ref{paramplane}.
	\label{phaseportraits}}
	\end{figure*}

Figure~\ref{phaseportraits} plots the phase portraits for the order parameter dynamics governed by (\ref{14}).  The panels show the qualitatively different behavior that occurs in the six regions of Fig.~\ref{paramplane}.  Figure~\ref{phaseportraits}(a) depicts what happens in the region where $b > 0$ and $\Omega < b$.  There, all trajectories are attracted to a stable fixed point on the unit circle, representing a synchronized rest state.  Notice that an invariant vertical line seems to join the repelling fixed point with the attracting one.  To prove that this vertical line truly is invariant, observe from (\ref{15}) that $\dot{x} = 0$ whenever $x = \Omega/b$.  Hence a solution that starts on this line stays there forever.

Figure~\ref{phaseportraits}(b) shows the case where $b > -1$ and $\Omega > |b|$.  The fixed points that previously existed on the unit circle have disappeared.  They annihilated each other when $\Omega = |b|$, thereby creating the periodic orbit on the unit circle seen in Fig.~\ref{phaseportraits}(b).  Physically, this orbit represents a synchronized oscillation with all the junctions moving in phase.  But this synchronous state is not attracting; it is neutrally stable.  In fact, the entire unit disk is filled with neutrally stable periodic orbits, all of which surround a neutrally stable fixed point, the splay state mentioned earlier.

The remaining panels show examples of additional cases when $b < 0$.  We do not dwell on these, as they correspond to a negative resistance in either the load or the Josephson junctions and hence are physically unrealistic.  The main features to observe are the saddle connection and the coexistence of two neutrally stable splay states in Fig.~\ref{phaseportraits}(c), and the saddle and center splay states in Fig.~\ref{phaseportraits}(f).
	
\section{$RLC$-LOADED CIRCUIT \label{RLC-LOADED CIRCUIT}}

\begin{figure}[ht]
	\includegraphics[scale = 1]{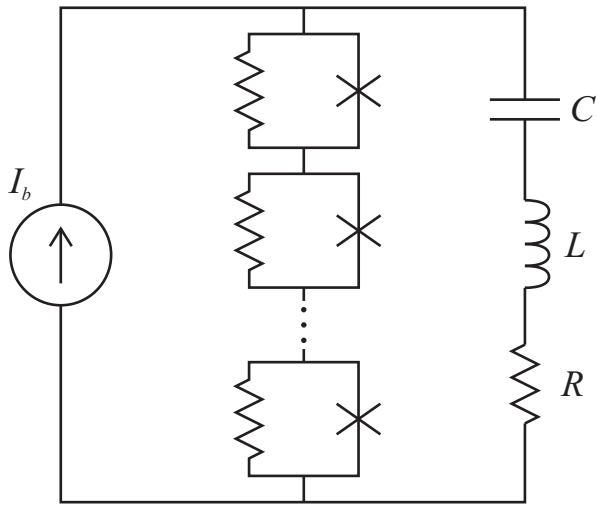}
	\caption{A series of Josephson junctions in parallel with an $RLC$ load.  $I_b$ represents a dimensionless version of the constant bias current, $C$ a dimensionless load capacitance, $L$ a dimensionless load inductance, and $R$ a dimensionless load resistance. \label{rlccircuit}}
	\end{figure}

We turn now to a more complicated kind of Josephson junction array.  Instead of the purely resistive load assumed earlier, we allow a load comprised of a resistor, inductor, and capacitor in series.  This load is placed in parallel with $N$ identical, overdamped Josephson junctions wired in series.  The whole circuit is driven by a constant current source $I_b$, as shown in Fig.~\ref{rlccircuit}.

Consider the infinite-$N$ limit of this system.  As shown by Strogatz and Mirollo~\cite{stro93}, the time-dependent dynamics of the array can be written in dimensionless form as
	\begin{equation} \label{28}
	v(\phi, t) = -\frac{i}{2}e^{-i\phi} + (I_b-\dot{Q}) + \frac{i}{2}e^{i\phi}
	\end{equation}
where the state variable $Q$ is governed by a dimensionless version of Kirchhoff's voltage law:
	\begin{equation} \label{29}
	L\ddot{Q} + (R+1)\dot{Q} + C^{-1}Q = I_b - \int_0^{2\pi} \rho(\phi, t) \sin{\phi} \mathrm{d}\phi
	\end{equation}
In (\ref{28}) and (\ref{29}), $Q(t)$ is the dimensionless charge on the capacitor, while the other new quantities are as indicated in Fig.~\ref{rlccircuit}.

Equation~(\ref{28}) has the special trigonometric form $v(\phi, t) = f e^{-i\phi} + g + \bar{f}e^{i\phi}$ required by the Ott-Antonsen method and therefore is reducible by the method of Sec.~\ref{REDUCIBLE SYSTEMS}.  By reading off the $f$ and $g$ implied by~(\ref{28}) and substituting them into the Riccati equation (\ref{riccati}) for the order parameter $z$, we obtain
	\begin{equation} \label{30}
	\dot{z} = \frac{1-z^2}{2} + i(I_b-\dot{Q})z,
	\end{equation}
which has real and imaginary parts
	\begin{equation} \label{31}
	\begin{split}
	\dot{r}    & = \frac{1-r^2}{2}\cos{\psi} \\
	\dot{\psi} & = \frac{r+r^{-1}}{2}\sin{\psi} + I_b - \dot{Q}.
	\end{split}
	\end{equation}
By defining $P = \dot{Q}$ and computing the integral, (\ref{29}) can also be split into a two-dimensional system:
	\begin{equation} \label{32}
	\begin{split}
	L\dot{P} & = -(R+1)P - C^{-1}Q + I_b - r\sin{\psi} \\
	\dot{Q}  & = P.
	\end{split}
	\end{equation}
Now let $P' = LP$, $Q' = LQ$, $S = -(R+1)/L$, and $\psi' = \pi/2-\psi$.  If we substitute these definitions into (\ref{31}) and (\ref{32}) and drop the primes, we obtain
	\begin{equation} \label{33}
	\begin{split}
	\dot{r}    & = \frac{1-r^2}{2}\sin{\psi}                    \\
	\dot{\psi} & = \frac{r+r^{-1}}{2}\cos{\psi} - I_b + L^{-1}P \\
	\dot{P}    & = I_b + SP - (LC)^{-1}Q - r\cos{\psi}          \\
	\dot{Q}    & = P.	
	\end{split}
	\end{equation}

This is the low-dimensional system that governs the flow on the Poisson submanifold. Observe that the state variables of (\ref{33}) are the order parameter amplitude $r$ and phase $\psi$, along with the dimensionless rescaled current $P$ and charge $Q$ through the load capacitor.  The control parameters are the bias current $I_b$, and the load inductance $L$, (rescaled) resistance $S$, and capacitance $C$.

A complete analysis of (\ref{33}) is beyond the scope of this paper.  From previous numerical experiments, we know that there would be many attractors and other complicated features to consider~\cite{tsan92,nich92,golo92,stro93}.  Rather than try to enumerate and analyze all of these, our aim will be to merely list a few basic facts about the fixed points of the system.  In particular, we show that an earlier result---an eigenvalue equation whose roots give the four non-trivial Floquet multipliers of the splay state---has a more straightforward derivation within the present framework.

\subsection{Fixed points \label{Fixed points 2}}

We first convert $r$ and $\psi$ of (\ref{33}) to Cartesian coordinates $x$ and $y$ as in the purely resistive case.  The result is
	\begin{equation} \label{34}
	\begin{split}
	\dot{x} & = -xy + (I_b-L^{-1}P)y                    \\
	\dot{y} & = \frac{1}{2}(x^2-y^2+1) - (I_b-L^{-1}P)x \\
	\dot{P} & = I_b + SP - (LC)^{-1}Q - x                     \\
	\dot{Q} & = P.	
	\end{split}
	\end{equation}
There are four fixed points of (\ref{34}):
	\begin{subequations} \label{35}
	\begin{align}
	x^* & = I_b;\; y^* =  \omega_-;\; P^* = 0;\; (LC)^{-1}Q^* = 0    \label{35a} \\
	x^* & = I_b;\; y^* = -\omega_-;\; P^* = 0;\; (LC)^{-1}Q^* = 0    \label{35b} \\
	x^* & = I_b + \omega_+;\; y^*,P^* = 0;\; Q^* =  \omega_+LC \label{35c} \\
	x^* & = I_b - \omega_+;\; y^*,P^* = 0;\; Q^* = -\omega_+LC \label{35d}	
	\end{align}
	\end{subequations}	
where $\omega_\pm = \sqrt{\pm I_b^2 \mp 1}$.  Once again, (\ref{35a}) and (\ref{35b}) represent synchronous fixed points in which the Josephson junctions are in phase and not oscillating, while (\ref{35c}) and (\ref{35d}) represent splay-state periodic orbits.  The derivation of (\ref{35}) makes no assumptions about the parameters except that they are real scalars, so we have no need to disregard certain parameter values as we did for $b$ in (\ref{16}).  Note that (\ref{35a}) and (\ref{35b}) exist if and only if $|I_b| \leq 1$.  Similarly, the requirement $|x^*| \leq 1$ implies (\ref{35c}) only exists on $(-\infty,-1]$ and (\ref{35d}) only exists on $[1,\infty)$.

\subsection{Stability of the splay state \label{Stability of the splay state}}

Since the splay states are the primary concern in much of the existing literature, we compute the linearization of (\ref{34}) at (\ref{35c}):
	\begin{equation} \label{38}
	\bm{J} = \left(
		\begin{array}{cccc}
		0           & \mp\omega_+ & 0                        & 0          \\
		\pm\omega_+ & 0           & L^{-1}(I_b \pm \omega_+) & 0          \\
		-1          & 0           & s                        & -(LC)^{-1} \\
		0           & 0           & 1                        & 0
		\end{array} \right)
	\end{equation}
The Jacobian (\ref{38}) has the characteristic polynomial:
	\begin{multline} \label{39}
	L\lambda^4 + (R+1)\lambda^3 + (C^{-1}+L\omega_+^2)\lambda^2 \\
	+ (I_b\omega_+ + R\omega_+^2)\lambda + \omega_+^2 C^{-1} = 0
	\end{multline}
where we have substituted back in the definition of $S$ and multiplied through by $L$.

The characteristic polynomial (\ref{39}) was first derived fifteen years ago (see (13) of Ref.~\cite{stro93}).  At the time, its derivation gave the first explanation for why there are just four non-neutral Floquet multipliers for the splay state of the $RLC$ system~\cite{nich92}.  It also allowed analytical predictions of those multipliers~\cite{stro93}.  However, the earlier derivation~\cite{stro93} involved Fourier expansions of infinitesimal perturbations about the splay states, a procedure more complicated than the one given here.

In retrospect, we can see now that perturbations tangent to the Poisson submanifold are precisely those responsible for the non-neutral directions; perturbations transverse to this manifold are the neutral ones.

\section{DISCUSSION \label{DISCUSSION}}

It is important to understand both the successes and the limitations of our analysis.  The systems we have studied comprise a special class of Josephson junction arrays, namely those in which all the junctions are identical and heavily overdamped, meaning that we can ignore their internal capacitance.  The junctions are connected in series, driven by a constant bias current and coupled through a load in parallel.  For this class of arrays, it has been known since 1994 that the governing equations specify a family of low-dimensional invariant manifolds~\cite{wata94}.  For resistively loaded arrays, these invariant manifolds are three-dimensional, while for arrays with an $RLC$ load, they are five-dimensional.  These results hold for any number of junctions and extend to infinite $N$.

In this paper, we have shown that in the infinite-$N$ limit, the equations can be reduced even more dramatically.  In other words, a strictly smaller invariant submanifold exists as a degenerate case of the invariant manifolds known previously.  We have called it the Poisson submanifold; it consists of all phase distributions taking the form of a Poisson kernel.  Using the Ott-Antonsen ansatz, we explicitly calculated the flow equations on this manifold, demonstrating in the process that the Poisson submanifold is invariant.

The resulting low-dimensional dynamical systems shed new light on the behavior of Josephson junction series arrays.  For example, earlier local arguments~\cite{tsan91,tsan92,nich92,golo92,stro93} showed that the synchronous periodic state and splay states for arrays with a resistive load exhibit neutral stability to linear order over a wide range of parameters.  But do they exhibit neutral stability in nonlinear reality?  This has been a long-standing open question.  We can now observe that the answer is yes. Figure~\ref{phaseportraits}(b) shows that the splay state is surrounded by neutrally stable periodic orbits that fill the Poisson submanifold, continuing all the way out to the synchronized orbit on the boundary of the disk.

The other phase portraits in Fig.~\ref{phaseportraits} similarly provide new information regarding the global structure.  For example, they elucidate how splay states bifurcate with one another and with the synchronous periodic state and reveal global features such as the vertical heteroclinic orbit joining the in-phase rest states in several panels of Figure~\ref{phaseportraits}.  Previous global results regarding Josephson junction arrays were confined to the averaged versions of such systems~\cite{wata93, wata94}, which were derived via perturbation methods in the limit of weak coupling or high bias current~\cite{swif92}.  Our results, by contrast, hold for all values of the circuit parameters.  The trade-off is that they require infinite $N$.

However, the most serious drawback of our analysis is that it focuses on a thin slice of phase space which is unrepresentative of the dynamics of the full system.  For instance, the Poisson submanifold for the resistively loaded array is a degenerate, two-dimensional leaf in the foliation of phase space by three-dimensional invariant manifolds.  Could attractors or other interesting dynamical states exist off the Poisson submanifold?  Numerical simulations say yes:  KAM-like chaos occurs in the original equations for resistively loaded arrays~\cite{golo92, wata94}.

The remaining challenge is then to show that the reduced equations faithfully capture this chaos on the larger invariant manifolds.  We will address this issue in a subsequent paper~\cite{miro09} in which we use M\"obius transformations to simplify the circuit equations for Josephson junction arrays.  Essentially the same idea has been developed independently by Pikovsky and Rosenblum~\cite{piko08}, who have obtained new results for the Kuramoto model as well as more complex hierarchies of oscillators.

\textbf{Acknowledgments:}
Research supported in part by National Science Foundation grant NSF CISE-0835706.

\end{document}